\documentclass[a4paper,11pt]{article}
\usepackage[utf8]{inputenc}
\usepackage[T1]{fontenc}
\usepackage{amsmath,amssymb,graphicx}
\makeatletter
\usepackage{babel}
\usepackage[active]{srcltx}
\usepackage{graphicx,color}
\usepackage{changebar}
\usepackage[T1]{fontenc}
\usepackage{esint}
\usepackage{multirow}
\usepackage{dsfont}
\usepackage{ae}
\usepackage{amsmath}
\usepackage{braket}
\usepackage{mathtools}
\usepackage{slashed}
\usepackage{empheq}
\usepackage{multirow}
\usepackage{graphicx}
\usepackage{amsfonts}
\usepackage{amsmath}
\usepackage{subfigure}

\pdfoutput=1 

\usepackage{jcappub} 

\usepackage[T1]{fontenc} 

\title{\boldmath Braneworlds in Warped Einsteinian  Cubic Gravity}

\author[a]{L. A. Lessa,}
\author[a,b]{R. V. Maluf}
\author[a,]{J. E. G. Silva}
\author[a,c]{C. A. S. Almeida}

\affiliation[a]{Universidade Federal do Cear\'a (UFC), Departamento de F\'isica,\\ Campus do Pici, Fortaleza - CE, C.P. 6030, 60455-760 - Brazil.}
\affiliation[b]{Departamento de F\'{i}sica Te\'{o}rica and IFIC, Centro Mixto Universidad de Valencia - CSIC. Universidad de Valencia, Burjassot-46100, Valencia, Spain.}
\affiliation[c]{Institute of Cosmology, Department of Physics and Astronomy, Tufts University, Medford, Massachusetts 02155, USA.}

\emailAdd{leandrolessa@fisica.ufc.br}
\emailAdd{r.v.maluf@fisica.ufc.br}
\emailAdd{euclides@fisica.ufc.br}
\emailAdd{carlos@fisica.ufc.br}

\abstract{

Einstenian cubic gravity (ECG) is a modified theory of gravity constructed with cubic contractions of the curvature tensor. This theory has the remarkable feature of having the same two propagating degrees of freedom of Einstein gravity (EG), at the perturbative level on maximally symmetric spacetimes. 
The additional unstable modes steaming from the higher order derivative dynamics are suppressed provided that we consider the ECG as an effective field theory wherein the cubic terms are seen as perturbative corrections of the Einstein-Hilbert term.
Extensions of ECG have been proposed in cosmology and compact objects in order to probe if this property holds in more general configurations. In this work, we construct a modified ECG gravity in a five dimensional warped braneworld scenario. By assuming a specific combination of the cubic parameters, we obtained modified gravity equations of motion with terms up to second-order. For a thin 3-brane, the cubic-gravity corrections yield an effective positive bulk cosmological constant. Thus, in order to keep the 5D bulk warped compact, an upper bound of the cubic parameter with respect to the bulk curvature was imposed. For a thick brane, the cubic-gravity terms modify the scalar field potential and its corresponding vacuum. Nonetheless, the domain-wall structure with a localized source is preserved.
At the perturbative level, the Kaluza-Klein (KK) tensor gravitational modes are stable and possess a localized massless mode provided the cubic corrections are small compared to the EG braneworld.

}

\begin{document}
\maketitle
\flushbottom

\section{Introduction}
High-curvature theories have assumed a pivotal role in recent years, particularly in the exploration of frameworks extending beyond General Relativity (GR). These extended models aim to tackle enduring challenges in large-scale physics, particularly phenomena like dark energy \cite{energy,SupernovaCosmologyProject:1998vns}, which seems to propel the expansion of the universe, while dark matter can explain the flatness of galactic rotation curves \cite{matter}. One prominent class of alternative high-derivative models is represented by $f(R)$ models, see Refs. \cite{fr1,fr2}. 
Moreover, it is conceivable to expand the $f(R)$ family by incorporating additional scalars, such as $R_{\mu\nu}R^{\mu\nu}$ and $R_{\mu\nu\alpha\beta}R^{\mu\nu\alpha\beta}$ . A prominent instance of this extension is the Gauss-Bonnet (GB) term, defined as $\mathcal{G}=R^2-4R_{\mu\nu}R^{\mu\nu} +R_{\mu\nu\alpha\beta}R^{\mu\nu\alpha\beta} $. An important characteristic of this curvature invariant is its absence of ghost instabilities \cite{Nunez:2004ts}. However, it's worth noting that this term becomes topological in $4$-dimensions. Nevertheless, there exist methodologies to introduce modifications to the equations of motion due to the GB term, even within the context of $d=4$, as explored in Ref. \cite{Glavan:2019inb}. Additionally, it is noteworthy that this term emerges in the low-energy string effective action \cite{Zwiebach:1985uq}. Recently, GB gravity has been applied to cosmology \cite{cosm1,cosm2}, black hole \cite{black1,black2} and branes \cite{gb1,gb2,gb}. 

In general, higher-order interactions are anticipated to manifest in the low-energy effective action, which serves as the UV completion of GR. Beyond the quadratic term of the GB invariant, there are also cubic scalar invariants in the curvature that emerge within the bosonic string framework \cite{string}. In recent times, a prominent category of gravitational theories incorporating high-curvature terms has gained considerable attention, known as Generalized Quasi-Topological Gravities (GQTGs). In the context of static spherically symmetric (SSS) black holes, as demonstrated in Ref. \cite{robie}, this theory exhibits second-order equations of motion. GQTGs can be further categorized into three distinct classes. The first encompasses the well-established Lovelock theory, recognized as the most encompassing metric theory of gravity that yields second-order field equations across various spacetime dimensions \cite{Lovelock:1971yv}. Notably, Lovelock theory is absent in four dimensions. The second category is characterized as Quasi-Topological Gravities (QTG), presenting a less restrictive framework compared to Lovelock's theory. The QTG, like Lovelock theory, also yields  equations of motion of second-order for SSS black hole solution \cite{Myers:2010ru}. Another intriguing characteristic of these two modified gravitational theories is that, in maximally symmetric backgrounds  (MSB), the linearized equations of motion align with Einstein's linearized equations, differing only by a general prefactor.

 Finally, the third case under consideration involves formulations incorporating up to cubic powers of the Riemann tensor. Remarkably, for $d=4$ and in the context of a SSS black hole solution, this theory yields second-order equations of motion, earning the designation of \textit{ Einsteinian Cubic gravity} (ECG) \cite{pablos}. It is noteworthy that, in contrast to the other two theories previously discussed, ECG is the most general cubic theory in $d=4$ which admits a single metric function spherical solution. In a broader context, this theory exhibits three key characteristics: (i) under the assumption of a MSB , it shares the spectrum with GR, signifying a massless and transverse graviton; (ii) the coefficients of each cubic invariant are dimension independent; (iii) notably, in four dimensions, the theory is neither trivial nor topological.  In the strong field regime, the ECG still produces singular black holes for the static solutions \cite{pablosp}. On the contrary, in Ref. \cite{Lessa:2023xto}, regular solutions featuring magnetic charge were discovered. Moreover, stationary \cite{spining1,spining2} and charged modified black hole solutions \cite{extremal,charged,charged2,charged3,spining3} were also found in ECG. In cosmology, the ECG dynamics can produce both an early and late expansion epochs \cite{cosmology1,cosmology2,cosmology3,cosmology4}, and with a bounce between two De Sitter vaccua \cite{bouncecosmology}. The ECG effects on quasinormal modes (QNM) \cite{qnm}, shadows \cite{shadow}
and gravitational lensing were also discussed \cite{lensing}.

Despite the remarkable properties of ECG reported above, the presence of higher-order terms lead to 
additional unstable modes \cite{ecginstability1}. Indeed, even on rather symmetric spacetimes, there is at least one ghost or unstable mode \cite{ecginstability1,ecginstability2}. The presence of such instabilities prevents the ECG from being considered as a fundamental theory. However, ECG can still be regarded as an effective field theory provided that the cubic terms be considered as perturbative corrections of the linear Einstein-Hilbert term \cite{ecginstability3,ecginstability4}.

Another noteworthy and unsolved challenge within GR is the hierarchy problem in the Standard Model of particle physics. A sophisticated resolution to this issue emerges when considering our universe as a 4-dimensional hypersurface embedded in a higher-dimensional spacetime, referred to as the \textit{bulk}. The Randall-Sundrum (RS) model \cite{brana1,brana2}, fitting into the broader context of braneworlds, stands out as a compelling framework that effectively addresses this high-dimensional theory. Nonetheless, various other models also delve into the exploration of the effects of extra dimensions, as detailed in Ref. \cite{brana3,brana4,brana5}. 
Furthermore, the thin brane solutions within the RS model, which exhibit singularities in the core, can be substituted with thick brane models. A notable example involves the consideration of thick branes generated by domain walls, where scalar fields serve as the source, as proposed in Ref. \cite{brana6, brana7}.

Inspired by the concept of domain walls in spacetime, a substantial body of literature has delved into the properties of thick branes, with a particular emphasis on gravitational aspects, see Refs.\cite{Costa:2015dva,Veras:2014bma,Dantas:2013iha}. Significantly, considerable progress has been achieved in the braneworld scenario through the implementation of the first-order formalism, known as Bogomolnyi-Prasad- Sommerfield (BPS) formalim \cite{bps1,bps2,bps3,bps4}, which has proven instrumental in generating stable thick brane solutions. Furthermore, the propagation of gravity and fields associated with mass and matter along the extra dimension can be scrutinized through the lens of their Kaluza-Klein spectrum. This analysis provides insights into how these fundamental forces extend into the additional spatial dimension, offering valuable perspectives within the framework of higher-dimensional theories.

The aim of this paper is to construct a modified ECG effective gravity theory in a five dimensional warped braneworld spacetime and to explore its properties.
We start by considering all eight possible action terms constructed from the contractions of the Riemann tensor up to cubic order. Even by imposing the same MSB spectrum condition used in Ref.\cite{pablos}, the EOM still have higher-order derivatives. Accordingly, we impose an additional condition on the cubic coefficients, leaving five independent coefficients. For a warped flat 3-brane metric, the modified ECG gravity EOM have terms up to second-order derivatives of the warp function. Considering a thin 3-brane solution, in order to obtain a warped configuration with a $AdS_5$ bulk spacetime, we imposed an upper bound for the cubic parameter $\lambda$. This condition defines a range of validity of this warped ECG effective theory, similar to those conditions found for ECG in four dimensional black holes geometries \cite{ecginstability3,ecginstability4}. For a thick 3-brane solution sourced by a minimally coupled scalar field, the warped ECG terms keep the brane stability as long as the effective field bound on the cubic parameter is satisfied.

In addition to the exact solutions of the warped ECG gravity, we also studied the propagation of the gravitational tensor modes on these modified braneworld geometries. It is worthwhile to mention that the braneworld solutions we considered are not MSB but asymptotically $AdS_5$. Thus, additional conditions on the cubic coefficients should be imposed in order to maintain the linearized gravitational equation on the curved bulk geometry up to second-order in derivatives. It turns out that the perturbed equation for the gravitational Kaluza-Klein (KK) modes depends only on a specific combination of the three remaining cubic coefficients. Incidentally, by imposing an upper bound on the cubic coupling constant $\lambda$, no tachyonic KK mode is allowed. 


The work is organized as the following. In section \ref{sec2} we construct an Einstenian cubic gravity in five dimensions. Since in $5D$ the Gauss-Bonnet term is no longer trivial, the five dimensional ECG action is constructed with the Einstein Hilbert term plus the quadratic (Gauss-Bonnet) and cubic (ECG) terms. 
In section \ref{sec3} we specialize the analysis for a five dimensional warped geometry, imposing a condition on the cubic coefficients in order to the gravitational equations to have terms up to second-order. Then, we study the effects of the quadratic and cubic term on thin 3-brane solutions in subsection \ref{sec3.1} and for thick brane solutions generated by scalar field using the first order formalism in the subsection \ref{sec3.2}. In section \ref{sec4}, we obtain the linearized equations for gravitational fluctuations up to cubic order explicitly. In section \ref{sec5}, we analyze the stability of the model and the problem of Kaluza-Klein (KK) mode localization.  Final remarks are summarized in section \ref{con}. Throughout the text, we adopt the capital Roman indices ($A,B,... = 0,1,2,3,4$) denote 5-dimensional bulk spacetime indices, the Greek indices ($\mu , \nu , ... = 0,1,2,3$) the spacetime indices of the worldbrane.
Moreover, we adopt the metric signature $(-,+,+,+)$.


\section{Einstenian cubic gravity in 5D} 
\label{sec2}

In this section, we construct an Einstenian cubic gravity (ECT) in a five dimensional curved spacetime as an effective field theory. Besides the usual Einstein-Hilbert (EH) term, we consider quadratic and cubic terms as perturbative corrections of the EH term. In five dimensions, the quadratic terms cannot be neglected, and under certain conditions, this interaction  is indeed the Gauss-Bonnet (GB) term.

We start by considering all the possible terms constructed with contractions of the Riemann tensor up to cubic order.
As shown in Ref. \cite{Myers:2010ru}, there are a total of 13 possible contractions, but only 10 persist when we discard the total derivative terms. We can streamline the possibilities by focusing solely on contractions with the curvature tensor, excluding the derivatives of the tensor and the scalar of Ricci. This reduces the options to only 8. For the sake of simplicity, this article adopts this more straightforward set.

The action of such a theory can be written as 
\begin{equation}
    S = \int d^{5}x\sqrt{-g_5}\bigg[\frac{1}{2\kappa_5}(R-2\Lambda_0) + \sum_{i=0}^{3}\alpha_{i}\mathcal{L}_{(i)} ^{(2)} + \lambda \sum_{i=0}^{10}\beta_{i}\mathcal{L}_{(i)} ^{(3)} + \mathcal{L}_m\bigg],
\end{equation}
where $\kappa_5 = 8 \pi G_{5}$ is the Newton's constant in five dimensions and $\Lambda_0$ is the cosmological constant. Further, the $\mathcal{L}_m$ is the matter Lagrangian. The $\mathcal{L} ^{(2)}$ are complete sets of quadratic invariants and $\alpha_i$ are coupling constants with mass dimension. On the other hand, we have that $\mathcal{L}^{(3)}$ composes the cubic invariants where $\lambda$ is a parameter with length dimension and $\beta_{i}$ are dimensionless coupling constants. These Lagrangians are chosen to be composed of the following invariants: 
\begin{equation}
    \mathcal{L} ^{(2)}=\left\{R^2, R_{MN}R^{MN}, R^{MNAB}R_{MNAB}\right\}
\end{equation}
\begin{align} \nonumber
   & \mathcal{L} ^{(3)}=\{R_{M}{}^{A}{}_{N}{}^{B}R_{A}{}^{C}{}_{B}{}^{D} R_{C}{}^{M}{}_{D}{}^{N},\  R_{MN}{}^{AB}R_{AB}{}^{CD}R_{CD}{}^{MN},\ R_{MNCD}R^{MNC}{}_{E}R^{DE}, \\
   &R_{MNCD}R^{MNCD}R,\  R_{MNCD}R^{MC}R^{ND},\ R_{M}^{N}R_{N}^{C}R_{C}^{D}R_{D}^{M},\  R_{MN}R^{MN}R,{} R^3\}
\end{align}

The main objective of this paper is to focus on analyzing the effects of the cubic interaction within the context of a five-dimensional spacetime. In this scenario, the more general cubic term $P$ is expressed as follows:
\begin{align}\nonumber \label{pp}
    &P = \beta_1 R_{M}{}^{A}{}_{N}{}^{B}R_{A}{}^{C}{}_{B}{}^{D} R_{C}{}^{M}{}_{D}{}^{N} + \beta_2R_{MN}{}^{AB}R_{AB}{}^{CD}R_{CD}{}^{MN} + \beta_3R_{MNCD}R^{MNC}{}_{E}R^{DE} \\ 
&+\beta_4R_{MNCD}R^{MNCD}R+\beta_5R_{MNCD}R^{MC}R^{ND}+\beta_6R_{M}^{N}R_{N}^{C}R_{C}^{M}+\beta_7R_{MN}R^{MN}R +\beta_8R^3.
\end{align}
Unlike Lovelock theory, which excludes contributions from space-times with $d<6$ in cubic interactions \cite{Lovelock:1971yv}, the mentioned term contributes to the equation of motion in $d=5$, as we will demonstrate later. Additionally, there are eight independent free parameters, reducible to six under the assumption that the spectrum mirrors that of GR. In simpler terms, the metric perturbation on a MSB conveys only a transverse and massless graviton. This condition is met by enforcing the prescribed parameter relationships\cite{pablos}.
\begin{equation}\label{b1}
    \beta_7 = \frac{1}{20} (3 \beta_1-24 \beta_2-20 \beta_3-80 \beta_4-7 \beta_5-12 \beta_6),
\end{equation}
\begin{equation}\label{b2}
    \beta_8 = \frac{1}{200} (-9 \beta_1+52 \beta_2+40 \beta_3+160 \beta_4+6 \beta_5+16 \beta_6).
\end{equation}
Remarkable, these conditions lead to the famous GB term for the quadratic term, i.e., $\mathcal{L} ^{(2)} = R^2 - 4R_{MN}R^{MN} + R_{MNAB}R^{MNAB}$ \cite{pablos}.

Thus, the action, encompassing curvature up to cubic order and sharing the same spectrum as GR in the 5-dimensional context (ECG), can be expressed as follows:
\begin{equation}
    S = \int d^{5}x\sqrt{-g_5}\bigg[\frac{1}{2\kappa_5}(R-2\Lambda_0) + \alpha\mathcal{L}_{GB} + \lambda P  +\mathcal{L}_m \bigg],
\end{equation}
where $\mathcal{L}^{(2)}=\mathcal{L}_{GB}$. Moreover, we assume as matter field the real scalar field in the form
\begin{equation}\label{materia}
    \mathcal{L}_m = -\frac{1}{2}g^{MN} \mathcal{D}_{M}\phi \mathcal{D}_{N}\phi - V(\phi),
\end{equation}
where $V$ is the scalar potential and $\mathcal{D}_{M}$ is the covariant derivative with respect to the metric $g_{MN}$. So that the field equations of this theory can be written as
\begin{equation}\label{eqq1}
    G_{MN} + \Lambda_0 g_{MN} = \kappa_{5} (T_{MN}+\alpha Q_{MN} + \lambda H_{MN}),
\end{equation}
\begin{equation}
    g^{MN}\mathcal{D}_{M}\mathcal{D}_{N}\phi - \frac{\partial V}{\partial \phi}=0,
\end{equation}
where $T_{MN}$ is the matter energy-momentum tensor given by 
\begin{equation}
    T_{MN} = g_{MN}\mathcal{L}_m + \partial_{M}\phi\partial_{N}\phi .
    \end{equation}
The Lanczos tensor $Q_{MN}$, which is attributed to the GB term,  is given by
\begin{equation}
    Q_{MN} =  g_{MN} \mathcal{L}_{GB} -4 R R_{MN} + 8 R_{MC}R^{C}{}_{N} + 8 R_{MCND}R^{CD} - 4 R_{MCDE}R_{N}^{CDE}
\end{equation}
and the contribution from the cubic term is given by
\begin{align} \nonumber
   H_{MN} &=  - \frac{2}{\sqrt{g_5}}\frac{\delta(\sqrt{-g_5}P)}{\delta g^{MN}}.
\end{align}
Due to the complexity of the terms generated, $H$ is explicitly provided in the appendix \ref{Appendix}.


\section{Warped Einstenian cubic gravity braneworld}
\label{sec3}

Once we introduced the modified higher-order gravitational action in 5 dimensions, let us now investigate how the cubic gravity modifies the warped braneworld models. Consider the flat 3-brane warped metric of form
\begin{equation} \label{warp}
    ds^2 = e^{2A(y)}\eta_{\mu\nu}dx^{\mu}dx^{\nu}+dy^2,
\end{equation}
where $A(y)$ is the warp factor that depend only on the extra-dimensional coordinate $y$ and $\eta_{\mu\nu}$ is the 3-brane flat Minkowski metric. 

In addition to the conditions (\ref{b1}) and (\ref{b2}), let us impose that the EOM in Eq. (\ref{eqq1}) has at most up to second-order derivatives terms even in a non-perturbative regime (full equations). As a result, we have to impose the additional condition
\begin{equation} \label{b3}
    \beta_6 = -\frac{2}{81}(3 \beta_1-84 \beta_2-55 \beta_3-320 \beta_4-27 \beta_5).
\end{equation}
Thus, by adopting the conditions (\ref{b1}), (\ref{b2}), and (\ref{b3}), the cubic scalar takes the simple form 
\begin{equation}
    P = - 4\beta A^{'4}(5 A^{'2}+6A^{''}), 
\end{equation}
where the parameter $\beta$ is defined by the following combination
\begin{equation}
    \beta = -3 \beta_1+12 \beta_2+8 \beta_3+40 \beta_4.
\end{equation}
The additional condition (\ref{b3}) makes this theory an extension of the original ECG. Since the condition (\ref{b3}) steams from the requirement of the full EOM to have terms up to second-order derivatives in a warped five dimensional spacetime, we named this extended ECG theory a warped ECG gravity (WECG). A similar extension of ECG was performed for isotropic cosmological models, the so-called cosmological ECG (CECG) \cite{cosmology1}.


By assuming the metric ansatz Eq.(\ref{warp}) and a scalar field $\phi=\phi(y)$ depending only on the extra dimension $y$, we obtain the following equations of motion
\begin{equation}\label{eq1}
    3A'' + 6A'^2 = -\frac{1}{2}\kappa_{5} (\phi^{'2}+2V) - \Lambda_0 + 24 \kappa_{5} \alpha A^{'2}(A^{'2} + A'') + 2\kappa_{5}\lambda\beta A^{'4}(2A^{'2} + 3A''),
\end{equation}
\begin{equation}\label{eq2}
    6A'^2 = \frac{1}{2}\kappa_{5}(\phi^{'2}-2V)-\Lambda_0 + 24\kappa_{5}\alpha A^{'4}+4\kappa_{5}\lambda\beta A^{'6},
\end{equation}
\begin{equation}\label{eq3}
    \phi''+ 4A'\phi'-\frac{\partial V}{\partial \phi}=0.
\end{equation}
The prime stands for derivative with respect to $y$. Note that both GB and warped ECG gravities have a similar form of the usual braneworld equations \cite{brana1}, except for the additional terms proportional to powers of $A'$ and $A''$. For $\beta=0$, we obtain the GB gravity braneworld, recently discussed in the Ref.\cite{gb}.  Moreover, for $\alpha=0$ and $\beta=0$, we recover the GR-based braneworld Einstein equations.



From the modified gravitational equations Eq.(\ref{eq1}) and Eq.(\ref{eq2}) we can define the effective energy densities
\begin{equation}
    \rho_{GB} =  24  \alpha A^{'2}(A^{'2} + A''),
\end{equation}
\begin{equation}
    \rho_{cubic}=  2\lambda\beta A^{'4}(2A^{'2} + 3A''),
\end{equation}
and the corresponding pressures
\begin{equation}
    p_{GB} =  -24\alpha A^{'4},
\end{equation}
\begin{equation}
    p_{cubic}=  -4\lambda\beta A^{'6}.
\end{equation}
It is worthwhile to mention that, for $\alpha>0$ and $\beta >0$, the higher curvature terms produce negative effective pressures, regardless the warp function solution $A(y)$.
By using the densities and pressures above, we can analyse how the GB and cubic gravities modify the GR-based braneworld solutions as if there were an additional source.



\subsection{Thin 3-brane}
\label{sec3.1}

Once we obtained the modified gravitational equations due to the warped ECG, let us now consider the corrections due to the
quadratic and cubic modification on a thin brane. This solution though simple enables us to find an upper bound for the warped cubic parameter $\lambda$ in order to keep a well-defined bulk geometry.

Let us consider a thin 3-brane in $AdS_{5}/Z_2$ vacuum bulk $(\Lambda_0 <0)$, i.e., the Randall-Sundrum (RS) model \cite{brana1}. Assuming a particular solution $A(y)=-c |y|$ and $\phi=0$, the solution of Eq. (\ref{eq2}) gives us:
\begin{equation}
\label{thinbrane}
    \Lambda_0 = -2 c^2 \left(3-12 \alpha  c^2 \kappa_{5}-2\lambda\beta  c^4 \kappa_{5} \right).
\end{equation}
In order to have a $Ads_5$ bulk solution, i.e., $\Lambda_0 <0$, 
we impose the following condition
\begin{equation}
\label{thinbrane2}
0 \leq \lambda\beta < \frac{3}{2 c^2}\bigg( \frac{1}{c^2 \kappa_{5}} - 4 \alpha \bigg).
\end{equation}
Since $R=-20 c^2$, then the strength of the cubic term $\lambda\beta$ decreases as the bulk curvature increases. Thus, the Eq.(\ref{thinbrane2}) provides an upper bound to the cubic coupling steaming from the extra dimensional model. Moreover, Eq.(\ref{thinbrane2}) defines the range of the warped ECG wherein this effective theory is well-behaved. Note that the upper bound of $\lambda\beta$ depends on the geometric invariant of the spacetime considered. Similar upper bounds for ECG in other spacetimes were found in cosmology \cite{ecginstability1} and black holes \cite{ecginstability3,ecginstability4}.

It is worthwhile to mention that both quadratic and cubic terms yield to positive bulk cosmological constant terms in Eq.(\ref{thinbrane}). Indeed, for the choice $A'=-c$, the effective cubic energy density is given by
\begin{equation}
\rho_{cubic}=-p_{cubic}=4\lambda\beta c^4, 
\end{equation}
whereas for the GB model,
\begin{equation}
\rho_{GB}=-p_{GB}=24\alpha c^4.    
\end{equation}
Accordingly, the higher order corrections to the RS model yields to $AdS_{5}-dS_5$ transition. Since a $dS_5$ is a GR solution with equation of state $p=-\rho$, then the quadratic and cubic corrections are equivalent to an effective source which violates the dominant energy condition $\rho \geq P$. Therefore, in order to keep a warped compactified solution, 
we have to impose that, for $\alpha=0$,
\begin{equation}
\label{thinbrane3}
0 \leq \frac{2}{3} \lambda\beta\kappa_5 c^4 < 1.
\end{equation}
Thus, the warped ECG correction should be considered as a perturbation on the GR braneworld solution to obtain a well-defined warped geometry leading to a finite effective $3+1$ action (warped compactification) \cite{brana2}. The interpretation of ECG and its extensions as effective field theories where the higher curvature terms are perturbative corrections on the Einstein-Hilbert term is well discussed in the literature \cite{ecginstability1,ecginstability2,ecginstability3,ecginstability4}.


\subsection{BPS solution with a bulk scalar field} 
\label{sec3.2}


Now let us investigate the modifications made by the cubic gravity on a thick 3-brane. Since we are interested in a localized source, let us first discuss the asymptotic behaviour.

Outside the brane core, we expect the field to reach a vacuum value, i.e., $\phi'=V=0$. In this regime, the EOM yields to
\begin{equation}
    3A''(1-2\kappa_5 \lambda\beta A'^4)=0.
\end{equation}
Assuming that $(1-2\kappa_5 \lambda\beta A'^4)>0$, then $A''=0$. Thus, outside the brane core, the warp factor behaves as in the thin brane solution, i.e., $A'=-c$. 
Likewise for the thin brane, we impose an upper bound for the warped ECG coupling in the form
\begin{equation}
\label{thickupperbound}
    0\leq  \lambda\beta < \frac{1}{2\kappa_5 c^4}.
\end{equation}
Note that the upper bound in Eq.(\ref{thickupperbound}) is even stringer than the one imposed for the thin 3-brane solution in Eq.(\ref{thinbrane2}).
Since asymptotically, $e^{A}\rightarrow e^{-c|y|}$, then as $\lambda\beta$ increases the exponential decay of the warp function decreases. Thus, the upper bound in Eq.(\ref{thickupperbound}) is imposed in order to guarantee an asymptotically $AdS_5$ geometry for the thick braneworld.

In order to examine the effects of the modified gravity on the brane core, let us employ the well-knwon BPS formalism. First, let us assume the relation between the warp factor and the known superpotential $W(\phi)$ in the form
\begin{equation}\label{pot}
A'(y) = - \frac{1}{3} \kappa_{5}  W(\phi).
\end{equation}
By subtracting Eq. (\ref{eq1}) from Eq. (\ref{eq2}) under the assumption of Eq. (\ref{pot}), we arrive at the expression:
\begin{equation}\label{phi}
    \phi '(y)= - \frac{1}{81} \left(-81+72 \Tilde{\alpha}   W^2+2\Tilde{\beta}  W^4   \right)W_{\phi },
\end{equation}
where $W_{\phi }=\frac{\partial W}{\partial \phi}$. The scalar potential can be derived by substituting Eq. (\ref{phi}) into Eq. (\ref{eq2}), yielding:
\begin{equation}\label{p}
    V=\frac{W_{\phi }^2 \left(2\Tilde{\beta} W^4+72 \Tilde{\alpha}   W^2-81\right)^2+36 \kappa_{5}  W^2 \left(2\Tilde{\beta} W^4+108 \Tilde{\alpha} W^2-243\right)}{13122},
\end{equation}
where $\Tilde{\alpha}=\alpha\kappa_{5}^3$ and $\Tilde{\beta}=\lambda\beta\kappa_{5}^5$. Hence, by doing so, we transform the pair of second-order linear equations (\ref{eq1})-(\ref{eq2}) into a set of two first-order equations (\ref{pot})-(\ref{phi}). Note that the cubic gravity leads to additional fourth and sixth powers of the superpotential in the potential.



\begin{figure}[!ht] 
 \includegraphics[height=5.5cm]{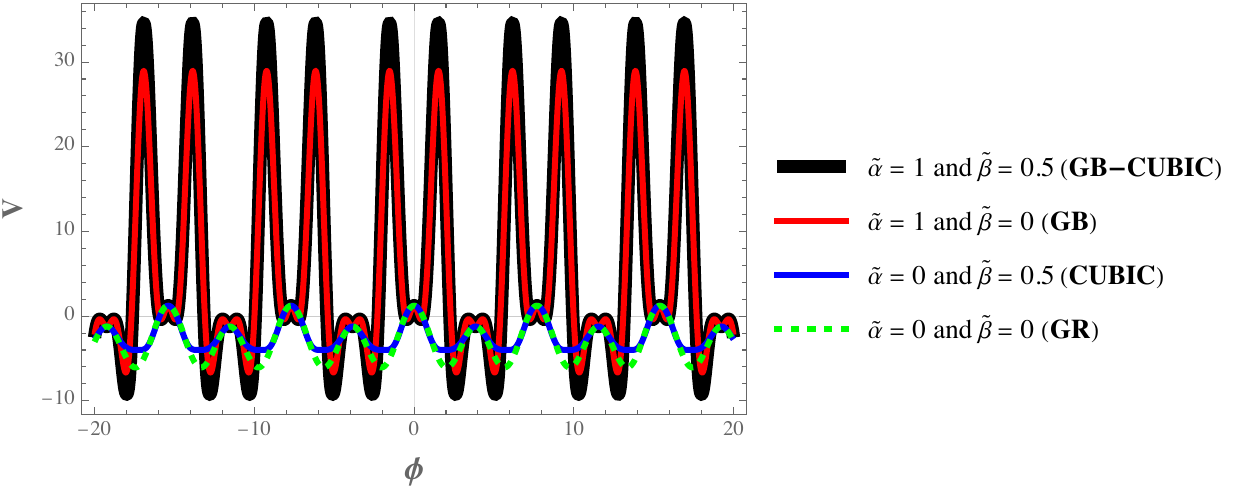}
\caption{ The potential profile given by Eq. (\ref{V}) for different values of $\Tilde{\alpha}$ and $\Tilde{\beta}$. We assume that $\kappa_{5}=1$ and $b=c=1$.}  \label{potential}
\end{figure}

\begin{figure}[h]

\center
\subfigure[ref1][The energy density]{\includegraphics[width=7.0cm]{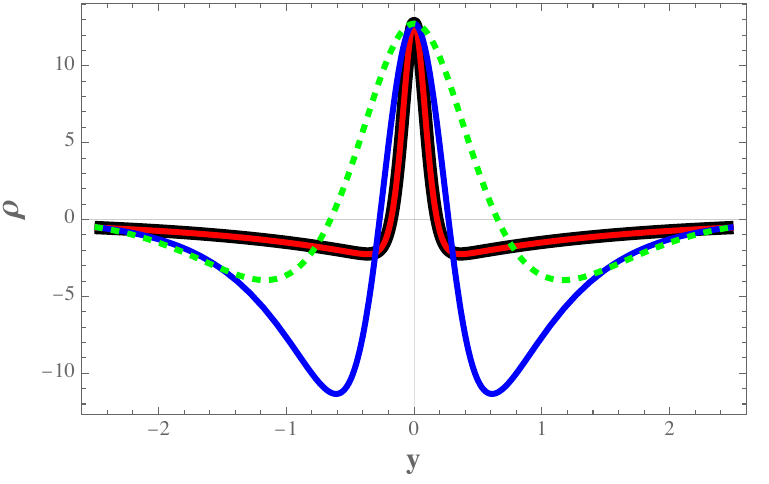}}
\qquad
\subfigure[ref2][The warped function]{\includegraphics[width=7.0cm]{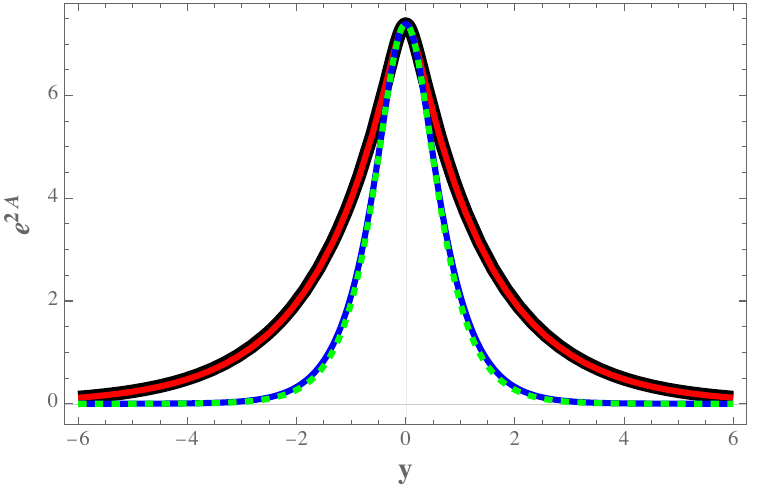}}
\qquad
\subfigure[ref2][The scalar field]{\includegraphics[width=11cm]{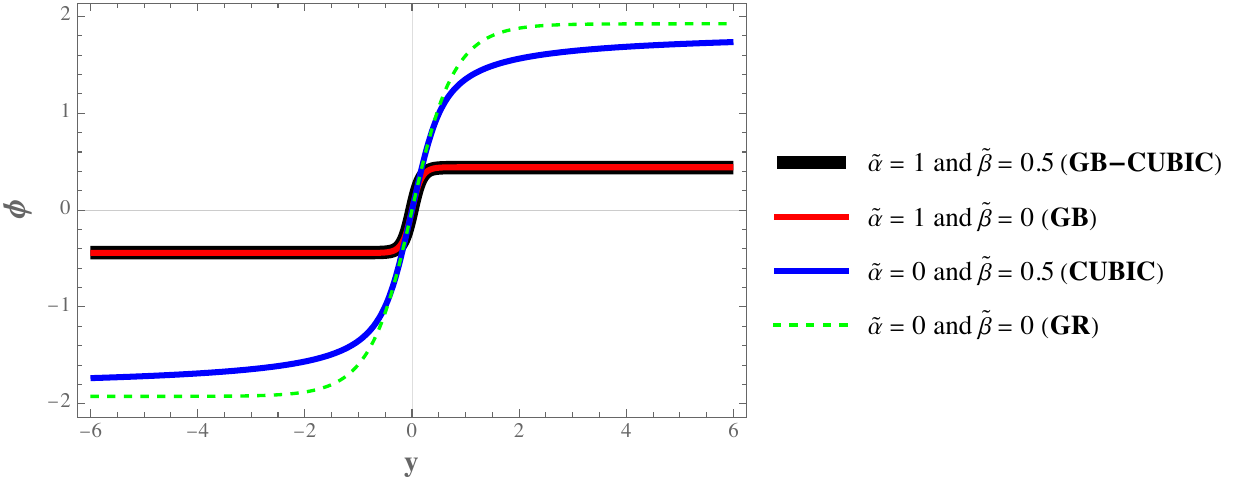}}
\caption{The trick brane solution for the superpotential $W=\xi \phi$. We assume that $\kappa_{5}=1$  and $\xi=1$.}
\label{fig8}
\end{figure}

Consider the well-known sine-Gordon model, where the superpotential is given by
\begin{equation}\label{super}
    W(\phi)=3 b c \sin \bigg[ \sqrt{\frac{2}{3b}}\phi \bigg],
\end{equation}
where $b$ and $c$ are constants. This straightforward selection yields an exact solution for the potential (\ref{p}), so that
\begin{align}\label{V} \nonumber
   &V=\frac{1}{2} c \bigg[4 b^2 c \kappa_5  \sin ^2\left(\sqrt{\frac{2}{3b}} \phi\right) \left(2 b^4 \Tilde{\beta}  c^4 \sin ^4\left(\sqrt{\frac{2}{3b}} \phi \right)+12 \Tilde{\alpha}  b^2 c^2 \sin ^2\left(\sqrt{\frac{2}{3b}} \phi \right)-3\right) \\
   &+\sqrt{\frac{6}{b}} \cos \left(\sqrt{\frac{2}{3b}} \phi\right) \left(2 b^4 \Tilde{\beta}  c^4 \sin ^4\left(\sqrt{\frac{2}{3b}} \phi\right)+8 \Tilde{\alpha}  b^2 c^2 \sin ^2\left(\sqrt{\frac{2}{3b}} \phi\right)-1\right)^2\bigg]
\end{align}
In Fig.\ref{potential}, we plot this potential for different configurations. Note that both the GB and the warped ECG gravity enhance the height of the barriers and the depth of the potential wells. A similar behavior was found in symmetric teleparallel braneworld \cite{fqbrane}. The energy density is sketched in the Fig.\ref{fig8}-a, the warp function is shown in Fig. \ref{fig8}-b
and the scalar field profile is ploted in Fig. \ref{fig8}-c. Note that for small values of the GB and WECG parameters, the warp factor, energy density and the scalar field profile exhibit the same profile as those found in a GR-based solutions.

\section{Tensor modes fluctuations}
\label{sec4}

In the last section, we extended the Einstenian cubic gravity for a five dimensional warped braneworld model. By assuming additional conditions on the ECG coefficients, we obtained gravitational equations of the motion containing up to second-order derivatives terms even in the non-perturbative level. Moreover, we found exact solution for a thin brane and a numerical solution for a thick brane. In both cases, we used the full EOM. In this section, we study the behaviour of the braneworld solutions in the perturbative level. We study the propagation of small perturbations upon the curved modified gravity braneworld found in the previous section.

Let us consider the tensor fluctuations upon the braneworld solution of form
\begin{equation}
    ds^2 = e^{2A(y)}[(\eta_{\mu\nu}+h_{\mu\nu})]+dy^2,
\end{equation}
where $h_{\mu\nu}(x^{\mu},y)$ is the tensor perturbation of the background metric (\ref{warp}). We will begin by explicitly deriving expressions for fluctuations in both the GB and WECG cases. Only after this analysis will we proceed to assume the transverse and traceless gauge conditions.

The nonvanishing components of the Ricci scalar, Ricci tensor, and
Riemann tensor are 

\begin{equation}
R=-4\left(2A''+5A'^{2}\right),
\end{equation}
\begin{equation}
R_{\mu\nu}=-e^{2A}\left(A''+4A'^{2}\right)\eta_{\mu\nu},
\end{equation}
\begin{equation}
R_{44}=-4\left(A''+A'^{2}\right),
\end{equation}
\begin{equation}
\label{riemann}
R_{\mu\nu\alpha\beta}=-e^{4A}A'^{2}\left(\eta_{\mu\alpha}\eta_{\nu\beta}-\eta_{\mu\beta}\eta_{\nu\alpha}\right),
\end{equation}
\begin{equation}
R_{\mu4\nu4}=-e^{2A}\left(A''+A'^{2}\right)\eta_{\mu\nu}.
\end{equation}
Note that, since we are considering a spacetime with a non-constant curvature tensor, our analysis is more general than one performed on maximally symmetric spacetimes (MSS) by Ref.\cite{pablos}. Nonetheless, the induced Riemann tensor on the brane in Eq.(\ref{riemann}) is rather similar to the expression for MSS, except for the factor $e^{4A}A'^{2}$ which varies along the extra dimension. For a thin brane geometry, the bulk is an MSS $AdS_5$, whereas for a thick 3-brane solution, the bulk is only asymptotically $AdS_5$.

The nonvanishing perturbation expansion for the Ricci scalar, Ricci
tensor and Riemann tensor are, respectively 
\begin{equation}
\delta R=e^{-2A}\left[\partial_{\mu}\partial_{\nu}h^{\mu\nu}-\Box^{(4)}h\right] -5A'h'-h'',
\end{equation}
\begin{align}
\delta R_{\mu\nu} & =\frac{1}{2}\left(\partial_{\mu}\partial_{\alpha}h_{\ \!\nu}^{\alpha}+\partial_{\nu}\partial_{\alpha}h_{\ \!\mu}^{\alpha}-\partial_{\mu}\partial_{\nu}h-\Box^{(4)}h_{\mu\nu}\right)\nonumber \\
 & -\frac{1}{2}e^{2A}\left[h''_{\mu\nu}+A'\left(h'\eta_{\mu\nu}+4h'_{\mu\nu}\right)+2\left(A''+4A'^{2}\right)h_{\mu\nu}\right],
\end{align}
\begin{equation}
\delta R_{\mu 4}=\frac{1}{2}\left(\partial^{\nu}h'_{\mu\nu}-\partial_{\mu}h' \right),
\end{equation}
\begin{equation}
\delta R_{44}=-\frac{1}{2}\left(h''+2A'h'\right),
\end{equation}
\begin{align}
\delta R_{\mu\nu\rho\delta} & =\frac{1}{2}e^{2A}\left[\partial_{\mu}\partial_{\delta}h_{\nu\rho}-\partial_{\nu}\partial_{\delta}h_{\mu\rho}-\partial_{\rho}\partial_{\mu}h_{\delta\nu}+\partial_{\rho}\partial_{\nu}h_{\delta\mu}+e^{2A}\left(2A'(\eta_{\mu\delta}h_{\nu\rho}-2\eta_{\nu\delta}h_{\mu\rho})\right.\right.\nonumber \\
 & \left.\left.+\eta_{\nu\rho}(2h_{\mu\delta}A'+h'_{\mu\delta})-\eta_{\mu\rho}(2h_{\nu\delta}A'+h'_{\nu\delta})-\eta_{\nu\delta}h'_{\mu\rho}+\eta_{\mu\delta}h'_{\nu\rho}\right)A'\right],
\end{align}
\begin{equation}
\delta R_{\mu\nu\rho4}=\frac{1}{2}e^{2A}\left[\partial_{\mu}h'_{\nu\rho}-\partial_{\nu}h'_{\mu\rho}\right],
\end{equation}
\begin{equation}
\delta R_{\mu4\rho4}=-\frac{1}{2}e^{2A}\left[2h_{\mu\rho}\left(A''+A'^{2}\right)+2A'h'_{\mu\rho}+h''_{\mu\rho}\right],
\end{equation}
where $h=h^{\mu}_{\mu}$ and $\Box^{(4)} = \eta^{\mu\nu}\partial_{\mu}\partial_{\nu}$ is the four-dimensional d’Alembert operator on the brane.

Taking the scalar field perturbation as $\phi=\bar{\phi}+\delta\phi$,
one finds the perturbation of the stress-energy tensor as
\begin{equation}
\delta T_{\mu\nu}=-e^{2A}\left(\frac{1}{2}h_{\mu\nu}\bar{\phi}'^{2}+h_{\mu\nu}V+\eta_{\mu\nu}\delta\phi'\bar{\phi}'+\eta_{\mu\nu}\delta\phi V_{\phi}\right),
\end{equation}
\begin{equation}
\delta T_{\mu4}=\delta\phi'\partial_{\mu}\bar{\phi}+\bar{\phi}'\partial_{\mu}\delta\phi,
\end{equation}
\begin{equation}
\delta T_{44}=\bar{\phi}'\delta\phi'-V_{\phi}\delta\phi,
\end{equation}
where $\bar{\phi}=\bar{\phi}(x^{\mu},y),$ and $\delta\phi=\delta\phi(x^{\mu},y)$
stands for the background scalar field and its perturbation, respectively.

\subsection{Gauss-Bonnet case}

Using the linearized quantities above, the linear perturbation for the Eq.
(\ref{eqq1}) (for $\lambda=0$) assuming the transverse,
traceless (TT) tensor perturbation, i.e., $\partial^{\mu}h_{\mu\nu}=h^{\mu}_{\mu}=0$ becomes
\begin{align}
h''_{\mu\nu}+4A'h'_{\mu\nu}+e^{-2A}\Box^{(4)}h_{\mu\nu}\\
-2\kappa_{5}\eta_{\mu\nu}\left(\delta\phi'\bar{\phi}'+\delta\phi V_{\phi}\right)\\
-8\alpha\kappa_{5}\left[A'^{2}h''_{\mu\nu}+2A'\left(2A'^{2}+A''\right)h'_{\mu\nu}+e^{-2A}\left(A''+A'^{2}\right)\Box^{(4)}h_{\mu\nu}\right]\\
-h_{\mu\nu}\left[6\left(A''+2A'^{2}\right)+\kappa_{5}\left(\phi'^{2}+2V\right)-48\alpha\kappa_{5} A'^{2}\left(A''+A'^{2}\right)\right] & =0.
\end{align}
Note that in the above expression, the second line involves only the
background scalar field and its perturbation. Besides, the fourth
line is vanishing since it is the EoM for zero order (\ref{eq1}), when we disregard $\rho_{cubic}$ . Thus, the linear
perturbation equation reduces to: 
\begin{equation}
\left[1-8\alpha\kappa_{5} A'^{2}\right]h''_{\mu\nu}+4A'\left[1-4\alpha\kappa_{5}\left(2A'^{2}+A''\right)\right]h'_{\mu\nu}+e^{-2A}\left[1-8\alpha\kappa_{5}\left(A''+A'^{2}\right)\right]\Box^{(4)}h_{\mu\nu}=0
\end{equation}
By defining $\zeta\equiv2\alpha\kappa_{5}$, we have fluctuation equation with the GB term reads
\begin{align} \label{modosgb}
\left[1-4\zeta A'^{2}\right]h''_{\mu\nu}+4A'\left[1-2\zeta\left(2A'^{2}+A''\right)\right]h'_{\mu\nu}\nonumber \\
+e^{-2A}\left[1-4\zeta\left(A''+A'^{2}\right)\right]\Box^{(4)}h_{\mu\nu} & =0.
\end{align}
The tensor perturbation Eq.(\ref{modosgb}) was obtained and studied in details in the Ref.\cite{gb}. We took great care in meticulously deriving the step-by-step procedure leading to Eq. (\ref{modosgb}). 

\subsection{Warped ECG Case}


After reviewing the linearization with the GB term, let us consider the fluctuations equations due to the cubic term. Since we are dealing with fluctuations upon a non maximally symmetric spacetime, we carefully obtained the linearized gravitational equation.

After cumbersome calculations, it turns out that, in order to keep the EOM for the fluctuations with terms up to second-order, we have to impose, in addition to the already imposed conditions (\ref{b1}),(\ref{b2}) and (\ref{b3}), the additional condition on the coefficient $\beta_4$ 
\begin{equation}
\label{beta4}
\beta_{4}=\frac{1}{32}\left(3\beta_{1}-12\beta_{2}-7\beta_{3}\right),
\end{equation}
and similarly for the $\beta_5$ coefficient
\begin{equation}
\beta_{5}=\frac{1}{9}\left(6\beta_{1}-60\beta_{2}-29\beta_{3}-64\beta_{4}\right).
\end{equation}
Remarkably, for the same $\beta_4$ in Eq.(\ref{beta4}), there are two other choices for $\beta_5$, namely $\beta_{5}=-\frac{1}{9}\left(3\beta_{1}+24\beta_{2}+8\beta_{3}-32\beta_{4}\right)
$ or $\beta_{5}=\frac{1}{9}\left(-12\beta_{1}+12\beta_{2}+13\beta_{3}+128\beta_{4}\right)$. 

Accordingly, by using the transverse, traceless (TT) gauge, the linearized EOM for the fluctuation becomes
\begin{align}
h''_{\mu\nu}+4A'h'_{\mu\nu}+e^{-2A}\Box^{(4)}h_{\mu\nu}\nonumber\\
-2\kappa_{5}\eta_{\mu\nu}\left(\delta\phi'\bar{\phi}'+\delta\phi V_{\phi}\right)\nonumber\\
-\frac{3}{2}\lambda\kappa_{5}\left(\beta_{1}-4\beta_{2}-\beta_{3}\right)A'^{2}\left[A'^{2}h''_{\mu\nu}+4A'\left(A''+A'^{2}\right)h'_{\mu\nu}+e^{-2A}\left(2A''+A'^{2}\right)\Box^{(4)}h_{\mu\nu}\right]\\
-h_{\mu\nu}\left[2\Lambda+6\left(A''+2A'^{2}\right)+\kappa_{5}\left(\phi'^{2}+2V\right)-3\lambda\kappa_{5}\left(\beta_{1}-4\beta_{2}-\beta_{3}\right)A'^{4}\left(3A''+2A'^{2}\right)\right] & =0.\nonumber
\end{align}
Note that in the above expression, the second line involves only the
background scalar field and its perturbation. Besides, the fourth
line is vanishing since it is the EoM for zero order (\ref{eq1}), when we disregard $\rho_{GB}$.
The linear perturbation equation reduces to
\begin{align}
\left[1-\frac{3}{2}\lambda\kappa_{5}\left(\beta_{1}-4\beta_{2}-\beta_{3}\right)A'^{4}\right]h''_{\mu\nu}\nonumber\\
+4A'\left[1-\frac{3}{2}\lambda\kappa_{5}\left(\beta_{1}-4\beta_{2}-\beta_{3}\right)A'^{2}\left(A''+A'^{2}\right)\right]h'_{\mu\nu}\nonumber\\
+e^{-2A}\left[1-\frac{3}{2}\lambda\kappa_{5}\left(\beta_{1}-4\beta_{2}-\beta_{3}\right)A'^{2}\left(2A''+A'^{2}\right)\right]\Box^{(4)}h_{\mu\nu} & =0
\end{align}
It is worthwhile to mention that the linear equation for the gravitational modes is modified by the cubic terms in a particular combination of the $\beta_1, \beta_2$, and $\beta_3$ coefficients. Thus, let us define the resulting cubic parameter as
\begin{equation}
     \epsilon\equiv\frac{3}{2}\lambda\kappa_{5}\left(\beta_{1}-4\beta_{2}-\beta_{3}\right).
\end{equation}

Then, the gravitational tensor fluctuations obey the equation
\begin{equation} \label{modos}
\left[1-\epsilon A'^{4}\right]h''_{\mu\nu}+4A'\left[1-\epsilon A'^{2}\left(A''+A'^{2}\right)\right]h'_{\mu\nu}+e^{-2A}\left[1-\epsilon A'^{2}\left(2A''+A'^{2}\right)\right]\Box^{(4)}h_{\mu\nu}=0,
\end{equation}
where $\Box^{(4)} = \eta^{\mu\nu}\partial_{\mu}\partial_{\nu}$.
Note that, for $\epsilon=0$, we recover the GR-based fluctuations equation. 

The signs of the coefficients in the terms $\Box^{(4)}h_{\mu\nu}$ and $h''$ determine the 
causal propagation of the tensor perturbations.
In order to the coefficient of the leading term in Eq.(\ref{modos}) be positive, then the cubic coupling constant $\epsilon$ must satisfy
\begin{equation}
\label{upperbound}
    0\leq \epsilon < \frac{1}{Max (A')^4}.
\end{equation}
For a cubic modified thin brane, where $A'=-c$, the cubic coupling constant satisfies
\begin{equation}
\label{fluctuationbound}
 0\leq \epsilon < \frac{1}{c^4}.   
\end{equation}
Therefore, the cubic modifications should be treated as small perturbations compared to the GR term. The bound limit obtained in the linearlized regime in Eq.(\ref{fluctuationbound}) agrees with the one found for the exact solution in Eq.(\ref{thinbrane3}).
Moreover, the upper bound in Eq.(\ref{upperbound}) avoids instabilities steaming from the negative values of the coefficient of the $h''_{\mu\nu}$ in Eq.(\ref{modos}). Thus, the interval for the WECG coupling in (\ref{fluctuationbound}) also defines the range of validity of the WECG gravity as an effective field theory, similar to bounds found in Ref.\cite{ecginstability1,ecginstability2,ecginstability3,ecginstability4}.

\section{Stability and localization of cubic gravity} \label{sec5}

After obtained the equation for the tensor modes fluctuation upon the 5 dimensional spacetime, i.e., Eq. (\ref{modos}), in this section we analyze the propagation of the fluctuations on the 3-brane and along the extra dimension. This analysis is important to ensure that the (3+1) effective action on the 3-brane is well-defined. Moreover, the propagation along the extra dimensions might lead to gravitational instabilities worth to be investigated.

Let us start by performing the usual Kaluza-Klein decomposition, where
\begin{equation}
\label{kkdecomp}
h_{\mu\nu}(x,y)= \chi_{\mu\nu}(x^{\mu})\varphi(y).    
\end{equation}
Substituting Eq.(\ref{kkdecomp}) into the fluctuation equation (\ref{modos}) yields to
\begin{equation}\label{modos1}
  (\Box^{(4)}  - m^2) \chi_{\mu\nu}(x^\mu)=0
\end{equation}
\begin{equation} \label{modos2}
B^2 \varphi''(y)+C\varphi'(y)+m^2 \varphi(y)=0.
\end{equation}
where
\begin{equation}\label{b}
    B^2 = \frac{e^{2A}[1-\epsilon A'^{4}]}{[1-\epsilon A'^{2}(2A''+A'^{2})]}
\end{equation}
and 
\begin{equation}\label{c}
    C = \frac{4A'e^{2A}\left[1-\epsilon A'^{2}\left(A''+A'^{2}\right)\right]}{[1-\epsilon A'^{2}(2A''+A'^{2})]}.
\end{equation}
The Eq. (\ref{modos1}) describes the (3+1) KK graviton with mass $m$ on the 3-brane, whereas the Eq.(\ref{modos2}) governs the graviton propagation along the extra dimension.

\subsection{KK spectrum stability}

The propagation of the tensor fluctuations along the 3-brane in Eq.(\ref{modos1}) has a mass term steaming from the extra dimension. This mass, known as the KK mass, can be obtained using the propagation along the extra dimension Eq.(\ref{modos2}). In order to analyse the stability of the KK spectrum, we employ the analogue supersymmetric quantum-mechanic approach.

Let us perform the change of coordinates
\begin{equation}
\label{coordinate}
z=\int B^{-1}(y) dy,     
\end{equation}
where $B(y)$ is given by Eq.(\ref{b}). As a result, 
the Eq.(\ref{modos2}) assumes the form
\begin{equation}
    \ddot{\varphi}(z) + E(z)  \dot{\varphi}(z)+m^2 \varphi(z)=0,
\end{equation}
where $E(z)= \frac{C - \dot{B}}{B}$ and the dot represents the derivative with respect to the $z$ coordinate.  Now assuming the following change on the wave function 
\begin{equation}
 \varphi(z) = e^{-\frac{1}{2}\int Edz} \Phi(z),    
\end{equation}
then the function $\Phi$ satisfies a Schr\"{o}dinger-like equation given by
\begin{equation} \label{susy}
    (-\partial_{z}^2+ V_{eff})\Phi = m^2\Phi,
\end{equation}
where the effective potential $V_{eff}$ is given by
 \begin{equation}
     V_{eff}(z) = \bigg( \frac{E}{2} \bigg)^2 + \partial_{z} \bigg( \frac{E}{2} \bigg).
 \end{equation}
The form of the effective potential shows that it possesses a supersymmetric analogue structure of quantum mechanics. This SUSY-like symmetry guarantees the stability analysis for the KK modes \cite{Dantas:2015dca}. 
Indeed, by defining a superpotential $U_g = -\frac{E}{2}$ and the SUSY-like Hamiltonian operator $Q^{\dagger} Q$, where
$ Q = \partial_{z} + U_g $, then the Schr\"{o}dinger-like equation Eq.(\ref{susy}) takes the form
\begin{equation} \label{eqq}
    Q^{\dagger} Q \Phi = m^2 \Phi. 
\end{equation}
Therefore, the mass eigenvalue is positive, $m^2\ge 0$, and thus, there is no tachyonic KK gravitational modes. Accordingly, we conclude that the cubic modifications preserves the graviton stability, at least at the linearized regime.


\subsection{Massive modes}

Once we determined the stability of the KK spectrum, 
now we explore the effects of cubic gravity on these massive KK modes. Unfortunately, the change of coordinate $z=z(y)$ in Eq.(\ref{coordinate}) cannot be obtained analytically for the thick brane solutions. Thus, we study the massive modes in the $y$ coordinate, using the Eq.(\ref{modos2}).

Let us start with the thin brane, i.e., where $A'=-c$ and $A''=0$. Accordingly, the Eq. (\ref{modos2}) yields to
 \begin{equation} \label{modos3}
 \varphi''(y)-4 c \varphi'(y)+m^2 e^{2 c |y|}\varphi(y)=0,
\end{equation}
which is virtually the same KK equation in the RS model. The solution are combinations of Bessel functions, which are not normalizable, i.e., not localized around the 3-brane. Therefore, the cubic gravity modifications lead to the same KK gravitational spectrum for a thin 3-brane in a $AdS_{5}$ bulk spacetime. This is an expect result, since the Einsteinian cubic terms were chosen in order to keep the GR spectrum over maximally symmetric spacetimes.

\begin{figure}[h] 
       \includegraphics[height=4.4cm]{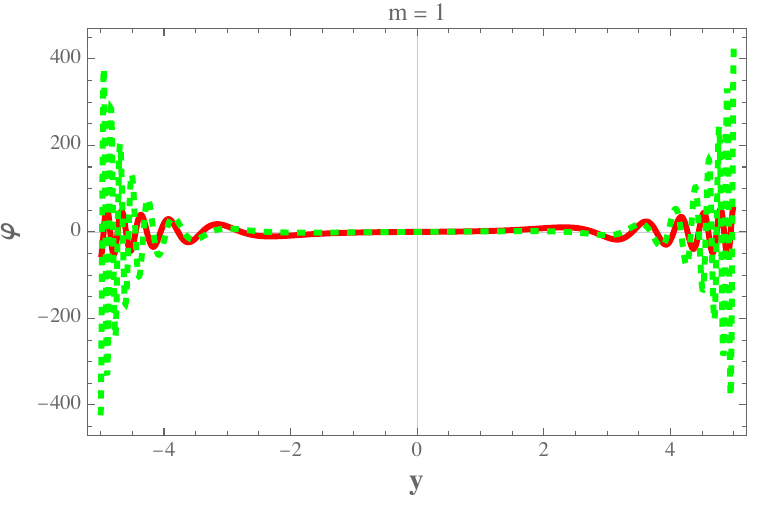}\quad
        \includegraphics[height=4.4cm]{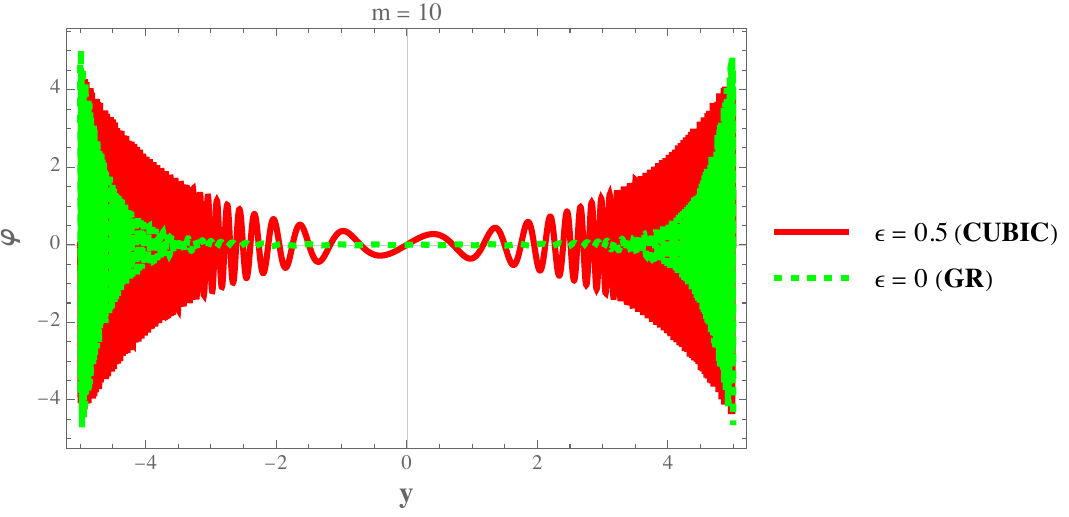}\quad
           \caption{Massive modes for light and heavy KK mass generated by a thick brane with (\ref{super}) superpotential. We assume that  $b=c=1$ .} 
          \label{fig81}
\end{figure}

For thick brane in the sine-Gordon model, we numerically solve the Eq.(\ref{modos2}). The behaviour of the KK massive modes is shown in the Fig. (\ref{fig81}). Note that, as the KK mass increases, the amplitude of the KK modes near the brane also increases.


\begin{figure}[h] 
       \includegraphics[height=4.4cm]{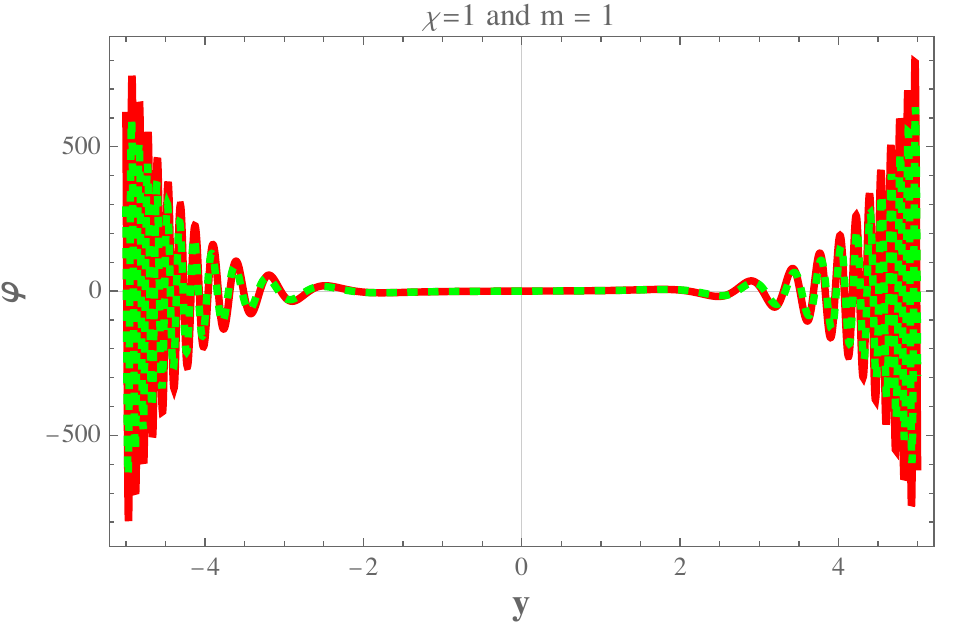}\quad
        \includegraphics[height=4.4cm]{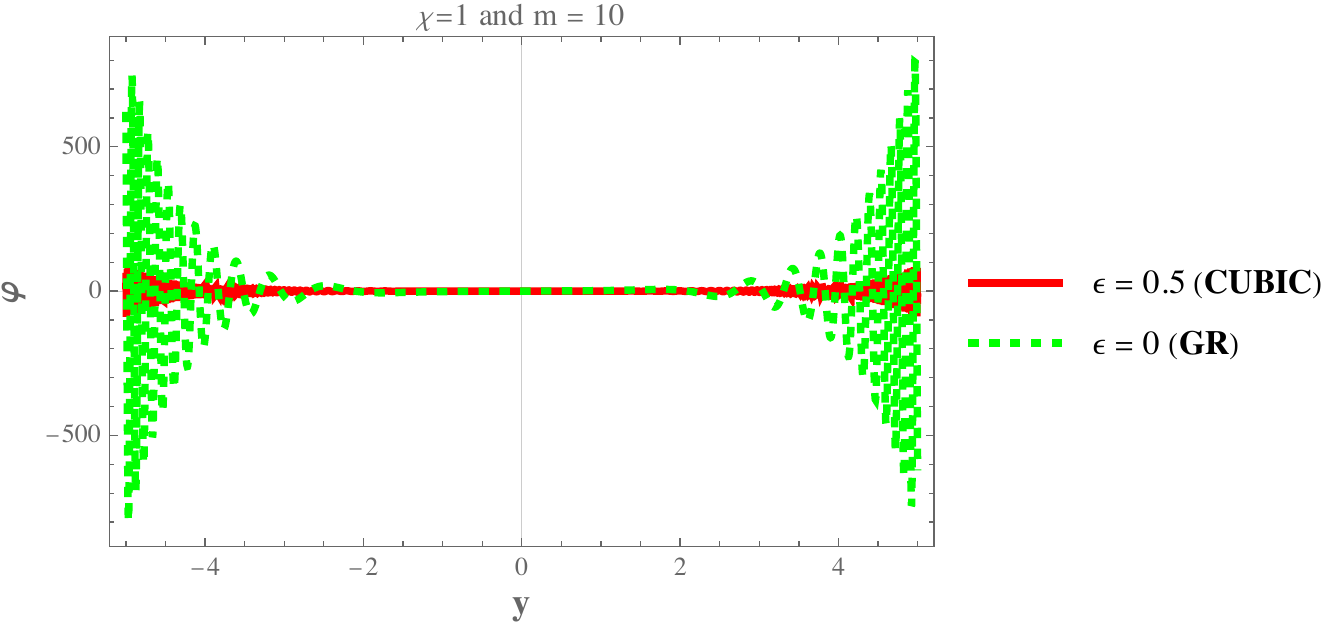}\quad
           \caption{Mass modes for light and heavy KK mass generated by a thick brane with $ A = \ln [sech(y)]$.} 
          \label{fig9}
\end{figure}

To counter this example, let's consider another thick brane now generated by a warped factor with the following ansatz
\begin{equation}\label{sec}
    A(y) = \ln [sech(\chi y)],
\end{equation}
where $\chi$ is a constant. Unlike the previous example, we now have the exact form of the warped factor, so we can substitute it into Eq. (\ref{b}) and (\ref{c}), so that
\begin{equation} \label{B1}
    B^2 = \frac{\text{sech}^2(\chi  y) \left(1-\chi ^4 \epsilon  \tanh ^4(\chi  y)\right)}{\chi ^4 \epsilon  \tanh ^4(\chi  y)-2 \chi ^4 \epsilon  \tanh ^2(\chi  y) \text{sech}^2(\chi  y)-1}
\end{equation}
\begin{equation}\label{c1}
    C = -\frac{4 \left(\chi  \tanh (\chi  y)-\chi^5 \epsilon  \tanh ^5(\chi  y)+\chi ^5 \epsilon  \tanh ^3(\chi  y) \text{sech}^2(\chi  y)\right)}{-\chi ^4 \epsilon  \tanh ^4(\chi  y)+2 \chi ^4 \epsilon  \tanh ^2(\chi  y) \text{sech}^2(\chi  y)+1}
\end{equation}
From this example with Eq. (\ref{b1}), it is already evident that the transformation $\frac{dy}{dz}=B$ used to obtain Eq. (\ref{susy}) may not lead to an invertible transformation between the coordinates $y$ and $z$. For this reason, we safely use Eq. (\ref{modos2}). Thus, numerically performing Eq. (\ref{modos2}) with (\ref{b1}) and (\ref{c1}), we construct the graphs in Fig. (\ref{fig9}). It's immediately noticeable that there is an inverse behavior compared to Fig. (\ref{fig8}), where for a heavy KK mass, the amplitudes of cubic gravity are suppressed by GR.

 \subsection{Massless modes}
Finally, let us consider the massless mode $m=0$. This mode must be confined to the brane to ensure that our theory leads to a finite action in $(3+1)$ dimensions. From Eq. (\ref{eqq}), the massless mode can be expressed as:
\begin{equation}
    \Phi_{0}(y) = \Tilde{\Phi}_0 e^{- \int U_{g}dz} 
\end{equation}
where $\Tilde{\Phi}_0$ is a normalization constant. Further, the localized graviton zero mode should satisfy the condition $\int_{-\infty} ^{-\infty} \Phi_0 ^2 dz=1$. Now, employing the $\frac{dy}{dz}= B(y)$ transformation, we proceed to numerically generate graphs depicting the zero modes as functions of the variable $y$. It is essential to highlight that the corresponding solution in this variable is expressed as follows

\begin{equation}
    \Phi_{0}(z) = \Tilde{\Phi}_0 e^{- \int U_{g}(y)dy} 
\end{equation}
where
\begin{equation}
    U_{g}(y) = - \frac{C - B'B}{2 B^2}.
\end{equation}

\begin{figure}[!ht] 

  \includegraphics[height=4.4cm]{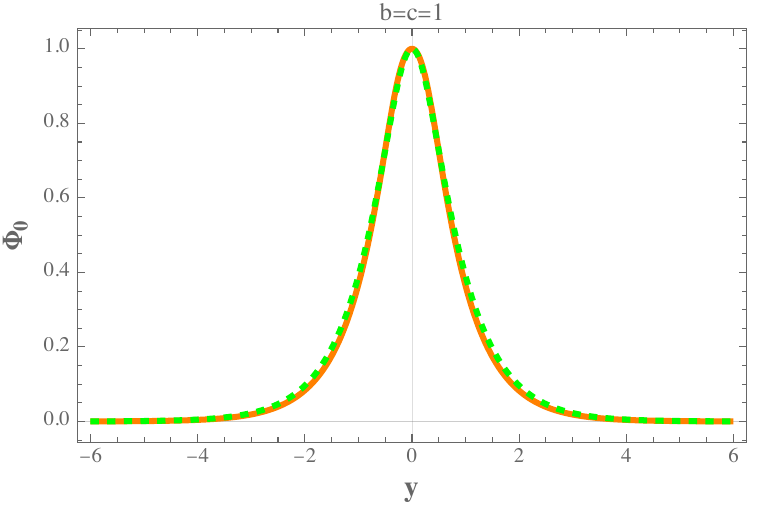}\quad
        \includegraphics[height=4.4cm]{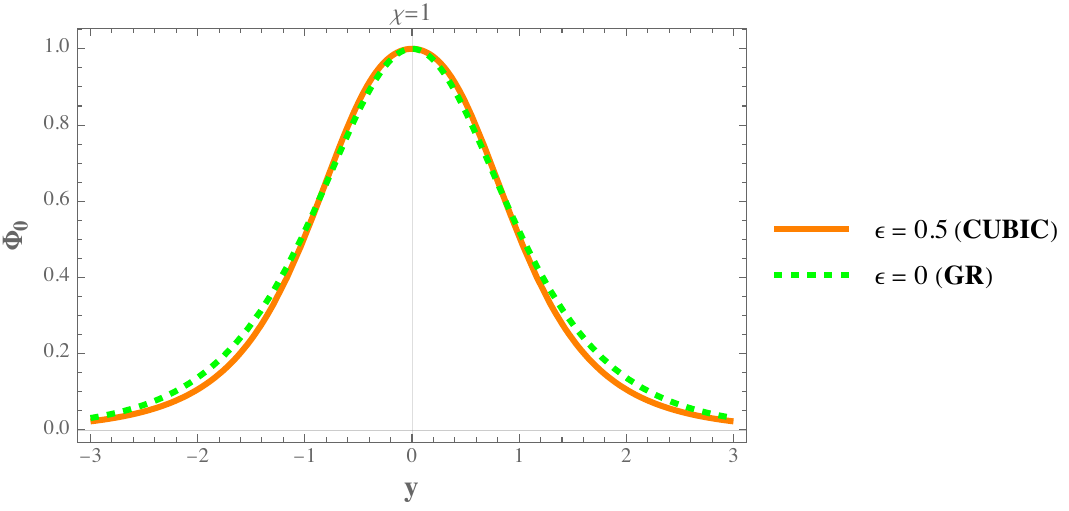}\quad
\caption{KK gravitational massless mode for weak cubic interaction. We assume that $\kappa_{5}=1$.}  \label{mz0}
\end{figure}

Considering the thick brane solution derived from the (\ref{super}) superpotential  and the  (\ref{sec}) warp factor, our analysis indicates that the zero mode exhibits localization at the origin, akin to the behavior observed in the standard RG case, refer to Figs. (\ref{mz0}). This phenomenon is particularly prominent under the assumption of a weak cubic interaction ($\epsilon=0.5$). 

\section{Final remarks}  \label{con}
In this work we investigated the effects of high curvature terms up to cubic order within the braneworld model. We proposed a cubic effective gravity theory inspired by Einsteinian Cubic Gravity (ECG), characterized by the invariant $P$ (\ref{pp}) determined by eight theory parameters $\beta$'s. Through the imposition that the theory shares a gravitational spectrum analogous to General Relativity (GR) in the linear regime, we successfully constrained two of these parameters. This specific parameterization not only defined the quartic term but also led to the complete specification of the Gauss-Bonnet (GB) term. Notably, at $d=5$, the GB term emerged as more than a topological term, justifying its inclusion in our calculations. Upon establishing a warped compactified brane geometry as defined by Eq. (\ref{warp}), it became evident that condition (i) alone was insufficient to yield second-order equations of motion for the warped factor. To address this limitation, an additional parameter needed to be fixed, so that we had to relax condition (ii). Consequently, through the fixation of three parameters, we successfully formulated a warped ECG gravity (WECG) featuring a second-order equation of motion for the brane model. Notably, this theory accurately describes the linear regime of the usual graviton for a maximally symmetric background (MSB). Additionally, we demonstrated that this theory gives rise to thin brane solutions reminiscent of the Randall-Sundrum (RS) type, exhibiting the potential for a de Sitter phase. For this brane, we have set an upper limit on the cubic parameter $\lambda$. This condition establishes the validity range of this warped ECG effective theory.

We introduced the first-order BPS formalism to derive thick brane solutions, incorporating a scalar field to represent the matter content in the bulk. Our investigation focused on solutions generated by the sine-Gordon superpotential, resulting in a potential (\ref{V}). Leveraging numerical methods, we successfully constructed the scalar field profile, revealing that the cubic theory supplemented by the GB term produces single-kink solutions. Additionally, when considering a weak quadratic-cubic interaction, the energy density profile closely resembles that of the standard GR.

Before delving into a thorough stability analysis of the model, we derived the equation governing the tensor gravitational perturbation explicitly. Our calculations for the quadratic interaction align with the previously predicted result in Ref.\cite{gb}. However, in the case of the cubic interaction, an extensive computation of terms revealed that fixing two additional parameters ($\beta_4$ and $\beta_5
$) is necessary to obtain second-order equations of motion for gravitational fluctuations. Consequently, achieving a cubic theory devoid of instability would require up to three free $\beta$ parameters. Notably, this analysis extends the initial premise of condition (i), which assumes that fluctuations occur in a SMB, as highlighted in \cite{pablos}. Finally, with these specified constraints and leveraging the analogous supersymmetric structure of quantum mechanics, we demonstrated the stability of our model. This implies the absence of tachyon Kaluza-Klein (KK) modes within our proposed framework. 

The initial example we investigated was the thin brane solution. We observed that the inclusion of cubic-curvature, assuming weak interaction, does not deviate significantly from the Kaluza-Klein (KK) modes observed in the standard case of GR. However, subtle modifications are observed in the massive KK tower for thick brane solutions. The amplitude of the ripples exhibits variations contingent on the mass parameter, leading to either an increase or decrease. Furthermore, we demonstrated that the massless four-dimensional tensor zero modes can be effectively localized on the thick branes, allowing for the recovery of the four-dimensional Newtonian potential.

\section*{Acknowledgments}
\hspace{0.5cm} The authors thank the Funda\c{c}\~{a}o Cearense de Apoio ao Desenvolvimento Cient\'{i}fico e Tecnol\'{o}gico (FUNCAP), the Coordena\c{c}\~{a}o de Aperfei\c{c}oamento de Pessoal de N\'{i}vel Superior (CAPES), and the Conselho Nacional de Desenvolvimento Cient\'{i}fico e Tecnol\'{o}gico (CNPq), Grant no. 200879/2022-7 (RVM) and Grand no. 304120/2021-9 (JEGS). 
R. V. Maluf thanks the Department of Theoretical Physics \& IFIC  of the University of Valencia - CSIC for the kind hospitality.
C. A. S. A. is supported by CNPq 309553/2021-0, CNPq/Produtividade. Furthermore, C. A. S. A. is grateful to the Print-UFC CAPES program, project number  88887.837980/2023-00, and acknowledges the Department of Physics and Astronomy at Tufts University for hospitality.

\appendix

\section{EoM to the warped Einstenian cubic gravity}: explicit results\label{Appendix}

In this appendix, we provide explicit expressions for the tensor $H_{MN}$ in the EoM (\ref{eqq1}) corresponding to the cubic gravity case, defined by
\begin{equation}
    H_{MN}=-\frac{2}{\sqrt{g_{5}}}\frac{\delta(\sqrt{-g_{5}}P)}{\delta g^{MN}},
\end{equation} where $P$ is given by Eq. (\ref{pp}). To enhance clarity, we break down the presentation into eight distinct pieces.
\begin{equation}
    H_{AB}\equiv H_{AB}^{(1)}+H_{AB}^{(2)}+H_{AB}^{(3)}+H_{AB}^{(4)}+H_{AB}^{(5)}+H_{AB}^{(6)}+H_{AB}^{(7)}+H_{AB}^{(8)}.
\end{equation}
	
To derive each term mentioned above, we employ the \textsc{xAct} suite of \textsc{Mathematica} packages for algebraic computations \cite{xact1}. The detailed results are outlined below:
\begin{align}
 & H_{AB}^{(1)}=\nonumber \\
 & \tfrac{1}{2}\beta_{1}\left\{ -3R^{FG}\bigl(4R_{A}{}^{H}{}_{B}{}^{J}R_{FHGJ}-(3R_{AFB}{}^{H}+R_{A}{}^{H}{}_{BF})R_{G}{}^{J}{}_{HJ}\bigr)+2\left[-g^{(5)}{}_{AB}R_{F}{}^{L}{}_{H}{}^{M}R^{FGHJ}R_{GLJM}\right.\right.\nonumber \\
 & +3R_{A}{}^{FGH}\left(3R_{BG}{}^{JL}R_{FJHL}+2R_{B}{}^{J}{}_{G}{}^{L}(R_{FJHL}+R_{FLHJ})+R_{B}{}^{J}{}_{F}{}^{L}R_{GJHL}-2R_{BGF}{}^{J}R_{H}{}^{L}{}_{JL}\right)\nonumber \\
 & -6\mathcal{D}_{A}R^{FG}\mathcal{D}_{B}R_{FG}+6\mathcal{D}_{B}R_{FG}\mathcal{D}^{G}R_{A}{}^{F}-6\mathcal{D}_{F}R_{BG}\mathcal{D}^{G}R_{A}{}^{F}+6\mathcal{D}_{A}R_{FG}\mathcal{D}^{G}R_{B}{}^{F}\nonumber \\
 & -3R_{AFBG}\mathcal{D}^{G}\mathcal{D}^{F}R+6R_{AFBG}\mathcal{D}_{H}\mathcal{D}^{H}R^{FG}-6\mathcal{D}_{G}R_{AFBH}\mathcal{D}^{H}R^{FG}-6\mathcal{D}_{G}R_{AHBF}\mathcal{D}^{H}R^{FG}\nonumber \\
 & +12\mathcal{D}_{H}R_{AFBG}\mathcal{D}^{H}R^{FG}+6R_{BFGH}\mathcal{D}^{H}\mathcal{D}_{A}R^{FG}+6R_{AFGH}\mathcal{D}^{H}\mathcal{D}_{B}R^{FG}-6R_{BFGH}\mathcal{D}^{H}\mathcal{D}^{G}R_{A}{}^{F}\nonumber \\
 & \left.\left.-6R_{AFGH}\mathcal{D}^{H}\mathcal{D}^{G}R_{B}{}^{F}+6R^{FGHJ}\mathcal{D}_{J}\mathcal{D}_{G}R_{AFBH}-6\mathcal{D}_{H}R_{BGFJ}\mathcal{D}^{J}R_{A}{}^{FGH}\right]\right\},
\end{align}

\begin{align}
 & H_{AB}^{(2)}=\nonumber \\
 & \beta_{2}\left\{ -24R^{FG}\left(R_{A}{}^{H}{}_{F}{}^{J}R_{BGHJ}+R_{AF}{}^{HJ}R_{BHGJ}\right)+24R_{B}{}^{F}R_{A}{}^{GHJ}R_{FHGJ}\right.\nonumber \\
 & +24R_{A}{}^{F}R_{B}{}^{GHJ}R_{FHGJ}+6R_{A}{}^{FGH}R_{BG}{}^{JL}R_{FHJL}+24R_{A}{}^{FGH}R_{B}{}^{J}{}_{G}{}^{L}R_{FJHL}\nonumber \\
 & -12R_{A}{}^{FGH}R_{B}{}^{J}{}_{GH}R_{F}{}^{L}{}_{JL}-6R_{A}{}^{FGH}R_{BF}{}^{JL}R_{GHJL}+6R_{A}{}^{FGH}R_{B}{}^{J}{}_{F}{}^{L}R_{GHJL}\nonumber \\
 & +g^{(5)}{}_{AB}R_{FG}{}^{LM}R^{FGHJ}R_{HJLM}+24\mathcal{D}_{F}R_{BG}\mathcal{D}^{G}R_{A}{}^{F}-24\mathcal{D}_{G}R_{BF}\mathcal{D}^{G}R_{A}{}^{F}\nonumber \\
 & \left.-24R_{BGFH}\mathcal{D}^{H}\mathcal{D}^{G}R_{A}{}^{F}-24R_{AGFH}\mathcal{D}^{H}\mathcal{D}^{G}R_{B}{}^{F}-12\mathcal{D}_{F}R_{BJGH}\mathcal{D}^{J}R_{A}{}^{FGH}\right\},
\end{align}

{\small
\begin{align}
 & H_{AB}^{(3)}=\nonumber \\
 & -\tfrac{1}{2}\beta_{3}\left\{ 4R^{FG}R_{AF}{}^{HJ}R_{BGHJ}+4R^{FG}R_{A}{}^{H}{}_{F}{}^{J}R_{BGHJ}+4R^{FG}R_{AF}{}^{HJ}R_{BHGJ}+8R^{FG}R_{A}{}^{H}{}_{F}{}^{J}R_{BHGJ}\right.\nonumber \\
 & -8R^{FG}R_{A}{}^{H}{}_{F}{}^{J}R_{BJGH}-8R^{FG}R_{A}{}^{H}{}_{B}{}^{J}R_{FHGJ}+2R_{A}{}^{FGH}R_{BG}{}^{JL}R_{FHJL}+8R_{A}{}^{FGH}R_{B}{}^{J}{}_{G}{}^{L}R_{FJHL}\nonumber \\
 & -2R_{A}{}^{FGH}R_{B}{}^{J}{}_{J}{}^{L}R_{FLGH}+2R_{A}{}^{F}{}_{F}{}^{G}R_{B}{}^{HJL}R_{GHJL}+2R_{A}{}^{FGH}R_{B}{}^{J}{}_{F}{}^{L}R_{GHJL}+4R^{FG}R_{AFB}{}^{H}R_{G}{}^{J}{}_{HJ}\nonumber \\
 & +4R^{FG}R_{A}{}^{H}{}_{BF}R_{G}{}^{J}{}_{HJ}-4g^{(5)}{}_{AB}R_{F}{}^{L}{}_{H}{}^{M}R^{FGHJ}R_{GLJM}-g^{(5)}{}_{AB}R_{FG}{}^{LM}R^{FGHJ}R_{HJLM}-2\mathcal{D}_{A}R_{BF}\mathcal{D}^{F}R\nonumber \\
 & -2\mathcal{D}_{B}R_{AF}\mathcal{D}^{F}R+4\mathcal{D}_{F}R_{AB}\mathcal{D}^{F}R-2R_{BF}\mathcal{D}^{F}\mathcal{D}_{A}R-2R_{AF}\mathcal{D}^{F}\mathcal{D}_{B}R-4R^{FG}\mathcal{D}_{G}\mathcal{D}_{A}R_{BF}\nonumber \\
 & +2R^{FGHJ}\mathcal{D}_{G}\mathcal{D}_{A}R_{BFHJ}-4R^{FG}\mathcal{D}_{G}\mathcal{D}_{B}R_{AF}+2R^{FGHJ}\mathcal{D}_{G}\mathcal{D}_{B}R_{AFHJ}+8R^{FG}\mathcal{D}_{G}\mathcal{D}_{F}R_{AB}\nonumber \\
 & +4R_{A}{}^{F}\left(R^{GH}R_{BGFH}+R_{F}{}^{G}R_{B}{}^{H}{}_{GH}-4R_{B}{}^{G}R_{F}{}^{H}{}_{GH}+\mathcal{D}_{G}\mathcal{D}^{G}R_{BF}\right)\\
 & +4R_{B}{}^{F}\left(R^{GH}R_{AGFH}+R_{F}{}^{G}R_{A}{}^{H}{}_{GH}+\mathcal{D}_{G}\mathcal{D}^{G}R_{AF}\right)-4\mathcal{D}_{B}R_{FG}\mathcal{D}^{G}R_{A}{}^{F}\nonumber \\
 & -8\mathcal{D}_{F}R_{BG}\mathcal{D}^{G}R_{A}{}^{F}+16\mathcal{D}_{G}R_{BF}\mathcal{D}^{G}R_{A}{}^{F}-4\mathcal{D}_{A}R_{FG}\mathcal{D}^{G}R_{B}{}^{F}+2R_{AFBG}\mathcal{D}^{G}\mathcal{D}^{F}R\nonumber \\
 & +2R_{AGBF}\mathcal{D}^{G}\mathcal{D}^{F}R+4\mathcal{D}_{A}R_{BFGH}\mathcal{D}^{H}R^{FG}+4\mathcal{D}_{B}R_{AFGH}\mathcal{D}^{H}R^{FG}+4\mathcal{D}_{G}R_{AFBH}\mathcal{D}^{H}R^{FG}\nonumber \\
 & +4\mathcal{D}_{G}R_{AHBF}\mathcal{D}^{H}R^{FG}+4R_{BFGH}\mathcal{D}^{H}\mathcal{D}_{A}R^{FG}+4R_{AFGH}\mathcal{D}^{H}\mathcal{D}_{B}R^{FG}+4R_{BFGH}\mathcal{D}^{H}\mathcal{D}^{G}R_{A}{}^{F}\nonumber \\
 & +4R_{BGFH}\mathcal{D}^{H}\mathcal{D}^{G}R_{A}{}^{F}+4R_{AFGH}\mathcal{D}^{H}\mathcal{D}^{G}R_{B}{}^{F}+4R_{AGFH}\mathcal{D}^{H}\mathcal{D}^{G}R_{B}{}^{F}\nonumber \\
 & +8g^{(5)}{}_{AB}R^{FGHJ}\mathcal{D}_{J}\mathcal{D}_{G}R_{F}{}^{L}{}_{HL}+2R_{B}{}^{FGH}\mathcal{D}_{J}\mathcal{D}^{J}R_{AFGH}+2R_{A}{}^{FGH}\mathcal{D}_{J}\mathcal{D}^{J}R_{BFGH}\nonumber \\
 & +2\mathcal{D}_{B}R_{FJGH}\mathcal{D}^{J}R_{A}{}^{FGH}+4\mathcal{D}_{J}R_{BFGH}\mathcal{D}^{J}R_{A}{}^{FGH}+2\mathcal{D}_{A}R_{FJGH}\mathcal{D}^{J}R_{B}{}^{FGH}\nonumber \\
 & \left.-4g^{(5)}{}_{AB}\mathcal{D}_{H}R_{G}{}^{L}{}_{JL}\mathcal{D}^{J}R^{FG}{}_{F}{}^{H}+4g^{(5)}{}_{AB}\mathcal{D}_{J}R_{G}{}^{L}{}_{HL}\mathcal{D}^{J}R^{FG}{}_{F}{}^{H}+g^{(5)}{}_{AB}\mathcal{D}_{L}R_{FGHJ}\mathcal{D}^{L}R^{FGHJ}\right\} ,\nonumber 
\end{align}
}

\begin{align}
 & H_{AB}^{(4)}=\nonumber \\
 & \beta_{4}\left\{ -8R^{FG}RR_{AFBG}+2R_{B}{}^{F}RR_{A}{}^{G}{}_{FG}-4RR_{A}{}^{FGH}R_{BFGH}+6R_{A}{}^{F}RR_{B}{}^{G}{}_{FG}\right.\nonumber \\
 & -2R_{AB}R_{FGHJ}R^{FGHJ}+g^{(5)}{}_{AB}RR_{FGHJ}R^{FGHJ}+16g^{(5)}{}_{AB}R_{F}{}^{L}{}_{H}{}^{M}R^{FGHJ}R_{GLJM}\nonumber \\
 & +4g^{(5)}{}_{AB}R_{FG}{}^{LM}R^{FGHJ}R_{HJLM}-8g^{(5)}{}_{AB}R^{FG}{}_{F}{}^{H}R_{G}{}^{JLM}R_{HJLM}+4\mathcal{D}_{A}R^{FGHJ}\mathcal{D}_{B}R_{FGHJ}\nonumber \\
 & +4R\mathcal{D}_{B}\mathcal{D}_{A}R+4R^{FGHJ}\mathcal{D}_{B}\mathcal{D}_{A}R_{FGHJ}-8R\mathcal{D}_{F}\mathcal{D}^{F}R_{AB}+8\mathcal{D}_{A}R_{BF}\mathcal{D}^{F}R\nonumber \\
 & +8\mathcal{D}_{B}R_{AF}\mathcal{D}^{F}R-16\mathcal{D}_{F}R_{AB}\mathcal{D}^{F}R-4R_{AFBG}\mathcal{D}^{G}\mathcal{D}^{F}R-4R_{AGBF}\mathcal{D}^{G}\mathcal{D}^{F}R\nonumber \\
 & \left.-16g^{(5)}{}_{AB}R^{FGHJ}\mathcal{D}_{J}\mathcal{D}_{G}R_{F}{}^{L}{}_{HL}-4g^{(5)}{}_{AB}\mathcal{D}_{L}R_{FGHJ}\mathcal{D}^{L}R^{FGHJ}\right\}, 
\end{align}

\begin{align}
 & H_{AB}^{(5)}=\nonumber \\
 & -\tfrac{1}{2}\beta_{5}\left\{ 8R_{F}{}^{H}R^{FG}R_{AGBH}+4R_{B}{}^{F}R^{GH}R_{AGFH}-4R^{FG}R_{A}{}^{H}{}_{H}{}^{J}R_{BFGJ}+4R_{A}{}^{F}R^{GH}R_{BGFH}\right.\nonumber \\
 & +4R^{FG}R_{A}{}^{H}{}_{F}{}^{J}R_{BGHJ}+4R^{FG}R_{AF}{}^{HJ}R_{BHGJ}-8R^{FG}R_{A}{}^{H}{}_{F}{}^{J}R_{BJGH}+4R^{FG}R_{AFG}{}^{H}R_{B}{}^{J}{}_{HJ}\nonumber \\
 & +8R^{FG}R_{A}{}^{H}{}_{B}{}^{J}R_{FHGJ}-4R_{A}{}^{F}R_{B}{}^{G}R_{F}{}^{H}{}_{GH}+4g^{(5)}{}_{AB}R^{FG}{}_{F}{}^{H}R_{G}{}^{J}{}_{J}{}^{L}R_{H}{}^{M}{}_{LM}\nonumber \\
 & +2g^{(5)}{}_{AB}R^{FG}{}_{F}{}^{H}R_{G}{}^{J}{}_{H}{}^{L}R_{J}{}^{M}{}_{LM}-8\mathcal{D}_{A}R^{FG}\mathcal{D}_{B}R_{FG}-\mathcal{D}_{A}R\mathcal{D}_{B}R-8R^{FG}\mathcal{D}_{B}\mathcal{D}_{A}R_{FG}\nonumber \\
 & +4\mathcal{D}_{F}R_{AB}\mathcal{D}^{F}R-2R_{BF}\mathcal{D}^{F}\mathcal{D}_{A}R-2R_{AF}\mathcal{D}^{F}\mathcal{D}_{B}R+4R^{FG}\mathcal{D}_{G}\mathcal{D}_{A}R_{BF}\nonumber \\
 & +4R^{FG}\mathcal{D}_{G}\mathcal{D}_{B}R_{AF}+4R^{FG}\mathcal{D}_{G}\mathcal{D}_{F}R_{AB}+4\mathcal{D}_{B}R_{FG}\mathcal{D}^{G}R_{A}{}^{F}-4\mathcal{D}_{F}R_{BG}\mathcal{D}^{G}R_{A}{}^{F}\nonumber \\
 & +4\mathcal{D}_{A}R_{FG}\mathcal{D}^{G}R_{B}{}^{F}-2g^{(5)}{}_{AB}R^{FG}{}_{F}{}^{H}\mathcal{D}_{H}\mathcal{D}_{G}R^{JL}{}_{JL}+4R_{AFBG}\mathcal{D}_{H}\mathcal{D}^{H}R^{FG}\nonumber \\
 & +4R^{FG}\mathcal{D}_{H}\mathcal{D}^{H}R_{AFBG}+2R_{AB}\mathcal{D}_{H}\mathcal{D}^{H}R^{FG}{}_{FG}+4\mathcal{D}_{A}R_{BFGH}\mathcal{D}^{H}R^{FG}\nonumber \\
 & +4\mathcal{D}_{B}R_{AFGH}\mathcal{D}^{H}R^{FG}+8\mathcal{D}_{H}R_{AFBG}\mathcal{D}^{H}R^{FG}+4R_{BFGH}\mathcal{D}^{H}\mathcal{D}_{A}R^{FG}\nonumber \\
 & +4R_{AFGH}\mathcal{D}^{H}\mathcal{D}_{B}R^{FG}+4g^{(5)}{}_{AB}R^{FGHJ}\mathcal{D}_{J}\mathcal{D}_{G}R_{F}{}^{L}{}_{HL}-8g^{(5)}{}_{AB}\mathcal{D}_{H}R_{G}{}^{L}{}_{JL}\mathcal{D}^{J}R^{FG}{}_{F}{}^{H}\nonumber \\
 & \left.+8g^{(5)}{}_{AB}\mathcal{D}_{J}R_{G}{}^{L}{}_{HL}\mathcal{D}^{J}R^{FG}{}_{F}{}^{H}+4g^{(5)}{}_{AB}R^{FG}{}_{F}{}^{H}\mathcal{D}_{L}\mathcal{D}^{L}R_{G}{}^{J}{}_{HJ})\right\},
\end{align}

\begin{align}
 & H_{AB}^{(6)}=\nonumber \\
 & \tfrac{1}{4}\beta_{6}\left\{ 8g^{(5)}{}_{AB}R^{FG}{}_{F}{}^{H}R_{G}{}^{J}{}_{J}{}^{L}R_{H}{}^{M}{}_{LM}+12g^{(5)}{}_{AB}R^{FG}{}_{F}{}^{H}R_{G}{}^{J}{}_{H}{}^{L}R_{J}{}^{M}{}_{LM}\right.\nonumber \\
 & +6\mathcal{D}_{A}R_{BF}\mathcal{D}^{F}R+6\mathcal{D}_{B}R_{AF}\mathcal{D}^{F}R+6R_{BF}\mathcal{D}^{F}\mathcal{D}_{A}R+6R_{AF}\mathcal{D}^{F}\mathcal{D}_{B}R\nonumber \\
 & +12R^{FG}\mathcal{D}_{G}\mathcal{D}_{A}R_{BF}+12R^{FG}\mathcal{D}_{G}\mathcal{D}_{B}R_{AF}+12\mathcal{D}_{A}R_{FG}\mathcal{D}^{G}R_{B}{}^{F}\nonumber \\
 & +12\mathcal{D}_{B}R_{FG}\mathcal{D}^{G}R_{A}{}^{F}-12R_{B}{}^{F}\left(R^{GH}R_{AGFH}-R_{F}{}^{G}R_{A}{}^{H}{}_{GH}+\mathcal{D}_{G}\mathcal{D}^{G}R_{AF}\right)\nonumber \\
 & -12R_{A}{}^{F}\left(2R_{B}{}^{G}R_{FG}+R^{GH}R_{BGFH}-R_{F}{}^{G}R_{B}{}^{H}{}_{GH}+\mathcal{D}_{G}\mathcal{D}^{G}R_{BF}\right)\nonumber \\
 & -24\mathcal{D}_{G}R_{BF}\mathcal{D}^{G}R_{A}{}^{F}-12g^{(5)}{}_{AB}R^{FG}{}_{F}{}^{H}\mathcal{D}_{H}\mathcal{D}_{G}R^{JL}{}_{JL}\nonumber \\
 & \left.-3g^{(5)}{}_{AB}\mathcal{D}_{H}R^{JL}{}_{JL}\mathcal{D}^{H}R^{FG}{}_{FG}-12g^{(5)}{}_{AB}\mathcal{D}_{H}R_{G}{}^{L}{}_{JL}\mathcal{D}^{J}R^{FG}{}_{F}{}^{H}\right\}, 
\end{align}

\begin{align}
 & H_{AB}^{(7)}=\nonumber \\
 & \beta_{7}\left\{ 2R_{B}{}^{F}RR_{A}{}^{G}{}_{FG}+2R_{A}{}^{F}R(-2R_{BF}+R_{B}{}^{G}{}_{FG})-2R_{AB}R^{FG}{}_{F}{}^{H}R_{G}{}^{J}{}_{HJ}\right.\nonumber \\
 & +g^{(5)}{}_{AB}RR^{FG}{}_{F}{}^{H}R_{G}{}^{J}{}_{HJ}+4\mathcal{D}_{A}R^{FG}\mathcal{D}_{B}R_{FG}+2\mathcal{D}_{A}R\mathcal{D}_{B}R\nonumber \\
 & +R^{FG}\left(-4RR_{AFBG}+4\mathcal{D}_{B}\mathcal{D}_{A}R_{FG}\right)+2R\mathcal{D}_{B}\mathcal{D}_{A}R-2R\mathcal{D}_{F}\mathcal{D}^{F}R_{AB}\nonumber \\
 & +2\mathcal{D}_{A}R_{BF}\mathcal{D}^{F}R+2\mathcal{D}_{B}R_{AF}\mathcal{D}^{F}R-4\mathcal{D}_{F}R_{AB}\mathcal{D}^{F}R-2R_{AB}\mathcal{D}_{H}\mathcal{D}^{H}R^{FG}{}_{FG}\nonumber \\
 & +2R_{BF}\mathcal{D}^{F}\mathcal{D}_{A}R+2R_{AF}\mathcal{D}^{F}\mathcal{D}_{B}R-2g^{(5)}{}_{AB}R^{FG}{}_{F}{}^{H}\mathcal{D}_{H}\mathcal{D}_{G}R^{JL}{}_{JL}\nonumber \\
 & -g^{(5)}{}_{AB}R\mathcal{D}_{H}\mathcal{D}^{H}R^{FG}{}_{FG}-2g^{(5)}{}_{AB}\mathcal{D}_{H}R^{JL}{}_{JL}\mathcal{D}^{H}R^{FG}{}_{FG}\nonumber \\
 & \left.-4g^{(5)}{}_{AB}\mathcal{D}_{J}R_{G}{}^{L}{}_{HL}\mathcal{D}^{J}R^{FG}{}_{F}{}^{H}-4g^{(5)}{}_{AB}R^{FG}{}_{F}{}^{H}\mathcal{D}_{L}\mathcal{D}^{L}R_{G}{}^{J}{}_{HJ}\right\}, 
\end{align}

\begin{align}
 & H_{AB}^{(8)}=\nonumber \\
 & \beta_{8}\left\{ 6R_{AB}R^{FG}{}_{FG}R^{HJ}{}_{HJ}-12\left(\mathcal{D}_{A}R^{F}{}_{F}\mathcal{D}_{B}R^{G}{}_{G}+R^{GH}{}_{GH}\mathcal{D}_{B}\mathcal{D}_{A}R^{F}{}_{F}\right)\right.\\
 & \left.+g^{(5)}{}_{AB}\left[12\mathcal{D}_{H}R^{JL}{}_{JL}\mathcal{D}^{H}R^{FG}{}_{FG}-R^{FG}{}_{FG}\left(R^{HJ}{}_{HJ}R^{LM}{}_{LM}-12\mathcal{D}_{L}\mathcal{D}^{L}R^{HJ}{}_{HJ}\right)\right]\right\} .\nonumber 
\end{align}


\end{document}